\documentclass[sigconf]{acmart}

\makeatletter
\def\subsubsection{\setlength\parindent{10pt}\@startsection{subsubsection}{3}%
  \z@{.5\linespacing\@plus.7\linespacing}{.1\linespacing}%
  {\normalfont\itshape}}
\makeatother

\usepackage{booktabs} 
\usepackage{array}
\usepackage[inline]{enumitem}
\usepackage{xspace}
\usepackage{graphicx}

\newlist{inlinelist}{enumerate*}{1}
\setlist*[inlinelist,1]{%
  label=(\roman*),
}
\author{Helia Hashemi}
\affiliation{%
  \institution{Center for Intelligent Information Retrieval}
  \institution{University of Massachusetts Amherst}
  \city{Amherst, MA 01003} 
}
\email{hhashemi@cs.umass.edu}

\author{Mohammad Aliannejadi}
\affiliation{%
  \institution{Faculty of Informatics}
  \institution{Universit{\`a} della Svizzera italiana (USI)}
  \city{Lugano} 
  \country{Switzerland}
}
\email{mohammad.alian.nejadi@usi.ch}

\author{Hamed Zamani}
\affiliation{%
  \institution{Center for Intelligent Information Retrieval}
  \institution{University of Massachusetts Amherst}
  \city{Amherst, MA 01003} 
}
\email{zamani@cs.umass.edu}

\author{W. Bruce Croft}
\affiliation{%
  \institution{Center for Intelligent Information Retrieval}
  \institution{University of Massachusetts Amherst}
  \city{Amherst, MA 01003} 
}
\email{croft@cs.umass.edu}

\copyrightyear{2019} 
\acmYear{2019} 
\setcopyright{acmcopyright}
\acmConference[arxiv '19]{arxiv '19}{}{}

\newcommand{\myparagraph}[1]{\vspace{0.3\baselineskip}\noindent{\textbf{#1}}.~}

\newcommand{\data}{ANTIQUE\xspace}

\begin{document}

\title{\data: A Non-Factoid Question Answering Benchmark}

\begin{abstract}

Considering the widespread use of mobile and voice search, answer passage retrieval for non-factoid questions plays a critical role in modern information retrieval systems. Despite the importance of the task, the community still feels the significant lack of large-scale non-factoid question answering collections with real questions and comprehensive relevance judgments. In this paper, we develop and release a collection of 2,626 open-domain non-factoid questions from a diverse set of categories. The dataset, called \data, contains 34,011 manual relevance annotations. The questions were asked by real users in a community question answering service, i.e., Yahoo! Answers. Relevance judgments for all the answers to each question were collected through crowdsourcing. To facilitate further research, we also include a brief analysis of the data as well as baseline results on both classical and recently developed neural IR models. 
\end{abstract}

\maketitle

\vspace{-0.2cm}
\section{Introduction}
\label{sec:intro}

With the rising popularity of information access through devices with small screens, e.g., smartphones, and voice-only interfaces, e.g., Amazon's Alexa and Google Home, there is a growing need to develop retrieval models that satisfy user information needs with sentence-level and passage-level answers. This has motivated researchers to study answer sentence and passage retrieval, in particular in response to \emph{non-factoid} questions~\cite{Cohen:2016, Yulianti:IEEE:2018}. Non-factoid questions are defined as open-ended questions that require complex answers, like descriptions, opinions, or explanations, which are mostly passage-level texts. Questions such as ``what is the reason for life?'' are categorized as non-factoid questions. 
We believe this type of questions plays a pivotal role in the overall quality of question answering systems, since their technologies are not as mature as those for factoid questions, which seek precise facts, such as ``At what age did Rossini stop writing opera?''.

Despite the widely-known importance of studying answer passage retrieval for non-factoid questions~\cite{Cohen:2016,Cohen:2018,Keikha:2014:SIGIR,Yulianti:IEEE:2018}, the research progress for this task is limited by the availability of high-quality public data. Some existing collections, e.g., \cite{Keikha:2014:SIGIR, Shah:2010:EPA:SIGIR}, consist of few queries, which are not sufficient to train sophisticated machine learning models for the task. Some others, e.g., ~\cite{Cohen:2016}, significantly suffer from incomplete judgments. Most recently, \citet{Cohen:2018:SIGIR} developed a publicly available collection for non-factoid question answering with a few thousands questions, which is called WikiPassageQA. 
Although WikiPassageQA is an invaluable contribution to the community, it does not cover all aspects of the non-factoid question answering task and has the following limitations: \begin{inlinelist}
\item it only contains an average of 1.7 relevant passages per questions and does not cover questions that have multiple aspects in multiple passages; 
\item it was created from the Wikipedia website, containing only formal text;
\item more importantly, the questions in the WikiPassageQA dataset were generated by crowdworkers, which is different from the questions that users ask in real-world systems; 
\item the relevant passages in WikiPassageQA contain the answer to the question in addition to some surrounding text. Therefore, some parts of a relevant passage may not answer any aspects of the question; 
\item it only provides binary relevance judgments.
\end{inlinelist}

To address these shortcomings, in this paper, we create a novel dataset for non-factoid question answering research, called \emph{\data},\footnote{\data stands for \underline{a}nswering \underline{n}on-fac\underline{t}o\underline{i}d \underline{que}stions.} with a total of 2,626 questions. In more detail, we focus on the non-factoid questions that have been asked by users of Yahoo! Answers, a community question answering (CQA) service. Non-factoid CQA data without relevance annotation has been previously used in \cite{Cohen:2016}, however, as mentioned by the authors, it significantly suffers from incomplete judgments.\footnote{More information on the existing collections is provided in Section~\ref{sec:rel}.} We collected a set of four-level relevance judgments through a careful crowdsourcing procedure involving multiple iterations and several automatic and manual quality checks. Note that we, in particular, paid extra attention to collect reliable and comprehensive relevance judgments for the test set. Therefore, we annotated the answers after conducting result pooling among several term-matching and neural retrieval models. In summary, \data provides annotations for 34,011 question-answer pairs, which is significantly larger than many comparable datasets. 

We further provide brief analysis to uncover the characteristics of \data. Moreover, we conduct extensive experiments with \data to present benchmark results of various methods, including classical and neural IR models on the created dataset, demonstrating the unique challenges \data introduces to the community. To foster research in this area, we release \data for research purposes.\footnote{\url{https://ciir.cs.umass.edu/downloads/Antique/}}

\vspace{-0.2cm}
\section{Existing Related Collections}
\label{sec:rel}

\myparagraph{Factoid QA Datasets} TREC QA~\cite{wang-smith-mitamura:2007:EMNLP-CoNLL2007} and WikiQA~\cite{Yang:2015:ACL} are examples of factoid QA datasets whose answers are typically brief and concise facts, such as named entities and numbers. InsuranceQA~\cite{Feng:2015:CoRR} is another factoid dataset in the domain of insurance. \data, on the other hand, consists of open-domain non-factoid questions that require explanatory answers. The answers to these questions are often passage level, which is contrary to the factoid QA datasets.

\myparagraph{Non-Factoid QA Datasets} There have been efforts for developing non-factoid question answering datasets~\cite{Habernal:SIGIR:2016,Keikha:2014:SIGIR,Yang:ECIR:2016}. \citet{Keikha:2014:SIGIR} introduced the WebAP dataset, which is a non-factoid QA dataset with 82 queries. The questions and answers in WebAP were not generated by real users. There exist a number of datasets that partially contain non-factoid questions and were collected from CQA websites, such as Yahoo! Webscope L6, Qatar Living~\cite{qatarliving}, and StackExchange. These datasets are often restricted to a specific domain, suffer from incomplete judgments, and/or do not contain sufficient non-factoid questions for training sophisticated machine learning models. The nfL6 dataset \cite{Cohen:2016} is a collection of non-factoid questions extracted from the Yahoo! Webscope L6. Its main drawback is the absence of complete relevance annotation. Previous work assumes that the only answer that the question writer has marked as correct is relevant, which is far from being realistic. That is why we aim to collect a complete set of relevance annotations. WikiPassageQA is another non-factoid QA dataset that has been recently created by \citet{Cohen:2018:SIGIR}. As mentioned in Section~\ref{sec:intro}, despite its great potentials, it has a number of limitations. \data addresses these limitations to provide a complementary benchmark for non-factoid question answering.\footnote{More information can be found in Section~\ref{sec:intro}.} More recently, Microsoft has released the MS MARCO V2.1 passage re-ranking dataset~\cite{Nguyen:2016:CoRR}, containing a large number of queries sampled from the Bing search engine. In addition to not being specific to non-factoid QA, it significantly suffers from incomplete judgments. In contrast, \data provides a reliable collection with complete relevance annotations for evaluating non-factoid QA models.


\myparagraph{Machine Reading Comprehension (MRC) Datasets} MRC has recently attracted a great deal of attention in the NLP community. The MRC task is often defined as selecting a specific short text span within a sentence, selecting the answer from predefined choices, or predicting a blanked-out word of a sentence. There exist a number of datasets for MRC, such as SQuAD~\cite{Rajpurkar:CoRR:2016}, BAbI~\cite{Weston:CoRR:2015}, and MS MARCO v1~\cite{Nguyen:2016:CoRR}. In this paper, we study retrieval-based QA tasks, thus MRC is out of the scope of the paper.




\begin{figure}[t]
    \centering
    \vspace{-0.4cm}
    \includegraphics[width=.96\linewidth]{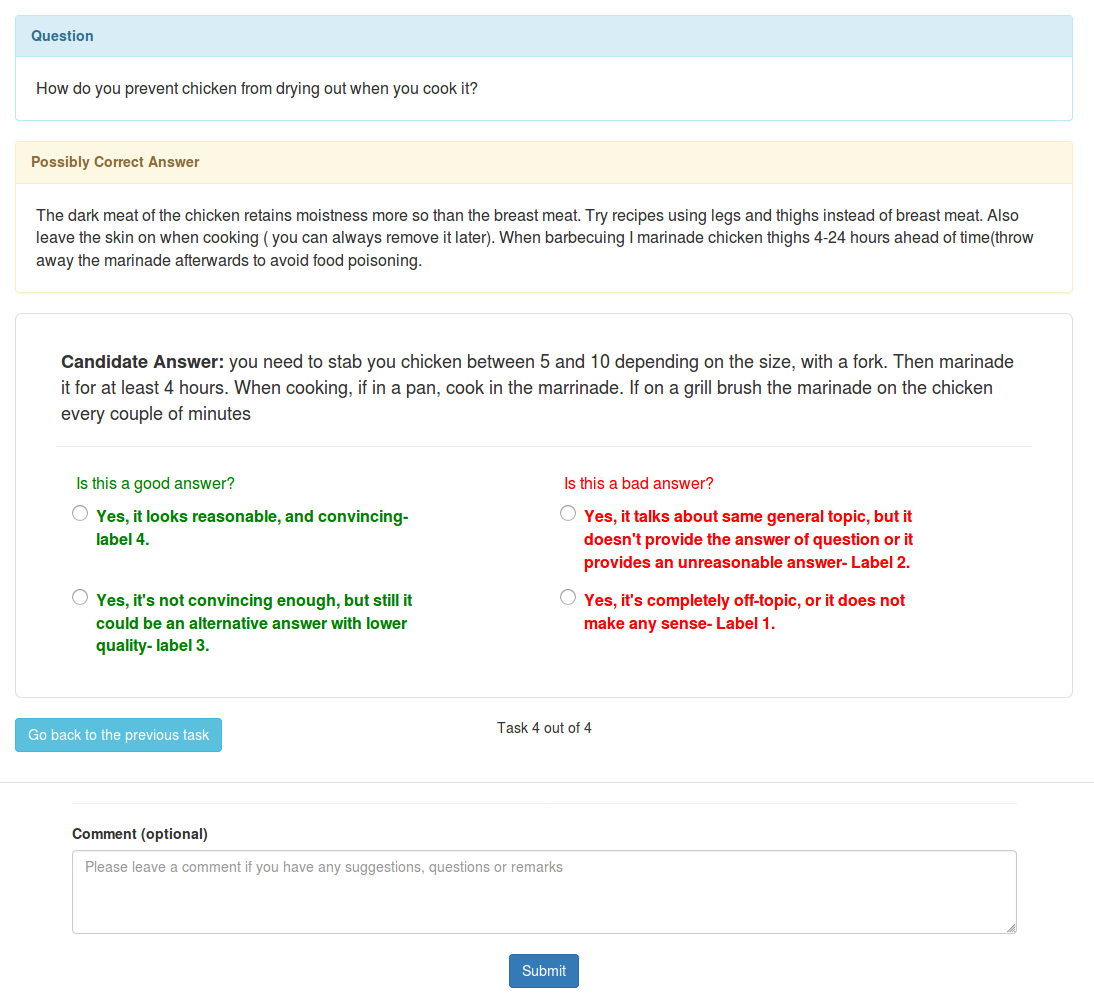}
    \caption{The HIT interface for answer relevance annotation.}
    \label{fig:worker_interface}
    \vspace{-0.6cm}
\end{figure}


\vspace{-0.2cm}
\section{Data Collection}
\label{sec:data}

In this section, we describe how we collected \data. Following \citet{Cohen:2016}, we used the publicly available dataset of non-factoid questions collected from the Yahoo! Webscope L6, called nfL6.\footnote{https://ciir.cs.umass.edu/downloads/nfL6/} 


\myparagraph{Pre-processing \& Filtering}
We conducted the following steps for pre-processing and question sampling: \begin{inlinelist}
\item questions with less than 3 terms were omitted (excluding punctuation marks);
\item questions with no best answer ($\hat{a}$) were removed;
\item duplicate or near-duplicate questions were removed. We calculated term overlap between questions and from the questions with more than 90\% term overlap, we only kept one, randomly;
\item we omitted the questions under the categories of ``Yahoo! Products'' and ``Computers \& Internet'' since they are beyond the knowledge and expertise of most workers;
\item From the remaining data, we randomly sampled 2,626 questions (out of 66,634).
\end{inlinelist}

Each question $q$ in nfL6 corresponds to a list of answers named ``nbest answers,'' which we denote with $\mathcal{A} = \{a_1, \dots, a_n\}$. For every question, one answer is marked by the question author on the community web site as the best answer, denoted by $\hat{a}$. 
It is important to note that as different people have different information needs, this answer is not necessarily the best answer to the question. Also, many relevant answers have been added after the user has chosen the correct answer. Nevertheless, in this work, we respect the users' explicit feedback, assuming that the candidates selected by the actual user are relevant to the query. Therefore, we do not collect relevance assessments for those answers.



\vspace{-0.2cm}
\subsection{Relevance Assessment}
We created a Human Intelligence Task (HIT) on Amazon Mechanical Turk,\footnote{\url{http://www.mturk.com/}} in which we presented workers with a question-answer pair, and instructed them to annotate the answer with a label between 1 to 4. The instructions started with a short introduction to the task and its motivations, followed by detailed annotation guidelines. Since workers needed background knowledge\footnote{Like for example ``Can someone explain the theory of $e=mc^2$?'' } for answering the majority of the questions, we also included $\hat{a}$ in the instructions and called it a ``possibly correct answer.'' Note that since we observed that, in some cases, the question was very subjective and could have multiple correct answers, we chose to call it a ``possibly correct answer'' and made it clear in the instructions that other answers could potentially be different from the provided answer, but still be correct. Figure~\ref{fig:worker_interface} shows the labeling interface where we provided a question and its ``possibly correct answer,'' asking workers to judge the relevance of a given answer to the question. 

\myparagraph{Label Definitions}
To facilitate the labeling procedure, we described the definition of labels in the form of a flowchart to users. Our aim was to preserve the notion of relevance in question answering systems as we discriminate it with the typical topical relevance definition in ad-hoc retrieval tasks. The definition of each label can be found in the following:
\begin{itemize}[leftmargin=*]
    \item \textit{Label 4}: It looks reasonable and convincing. Its quality is on par with or better than the ``Possibly Correct Answer''. Note that it does not have to provide the same answer as the ``Possibly Correct Answer''.
    \item \textit{Label 3}: It can be an answer to the question, however, it is not sufficiently convincing. There should be an answer with much better quality for the question.
    \item \textit{Label 2}: It does not answer the question or if it does, it provides an unreasonable answer, however, it is not out of context. Therefore, you cannot accept it as an answer to the question.
    \item \textit{Label 1:} It is completely out of context or does not make any sense.
\end{itemize}

Finally, we included 15 diverse examples of QA pairs with their annotations and explanation of why and how the annotations were done.


Overall, we launched 7 assignment batches, appointing 3 workers to each QA pair. In cases where the workers could agree on a label (i.e., majority vote), we considered the label as the ground truth. We then added all QA pairs with no agreement to a new batch and performed a second round of annotation.
It is interesting to note that the ratio of pairs with no agreement was nearly the same among the 7 batches (\textasciitilde13\%). 
In the very rare cases of no agreement after two rounds of annotation (776 pairs), an expert annotator decided on the final label.
To allow further analysis, we have added a flag in the dataset identifying the answers annotated by the expert annotator.
In total, the annotation task costed 2,400 USD.

\myparagraph{Quality Check}
To ensure the quality of the data, we limited the HIT to the workers with over 98\% approval rate, whi have completed at least 5,000 assignments.\footnote{We increased the previous assignment limit to 10,000 for annotating the test set.} 3\% of QA pairs where selected from a set of quality check questions with obviously objective labels. It enabled us to identify workers who did not provide high-quality labels. Moreover, we recorded the click log of the workers to detect any abnormal behavior (e.g., employing automatic labeling scripts) that would affect the quality of the data. Finally, we constantly performed manual quality checks by reading the QA pairs and their respective labels. The manual inspection was done on the 20\% of each worker's submission as well as the QA pairs with no agreement. 


\begin{table}[t]
    \centering
    \caption{Statistics of \data. 
    }
    \vspace{-0.3cm}
    \label{tab:stats}
    \begin{tabular}{ll}
        \toprule
         \# training (test) questions  & 2,426 (200) \\
         \# training (test) answers & 27,422 (6,589)\\
         \# terms/question & 10.51\\
         \# terms/answer & 47.75\\
         \midrule
         \# label 4  & 13,067 \\
         \# label 3  & 9,276 \\
         \# label 2  & 8,754 \\
         \# label 1  & 2,914 \\
         \midrule
         \# total workers & \multicolumn{1}{l}{577}\\
         \# total judgments & \multicolumn{1}{l}{148,252}\\
         \# rejected judgments & \multicolumn{1}{l}{17,460}\\
         \% of rejections & \multicolumn{1}{l}{12\%}\\
         \bottomrule
    \end{tabular}
    \vspace{-0.3cm}
\end{table}

\vspace{-0.2cm}
\subsection{Data Splits}
\label{sec:train set}


\begin{figure}
    \centering
    \includegraphics[width=\columnwidth]{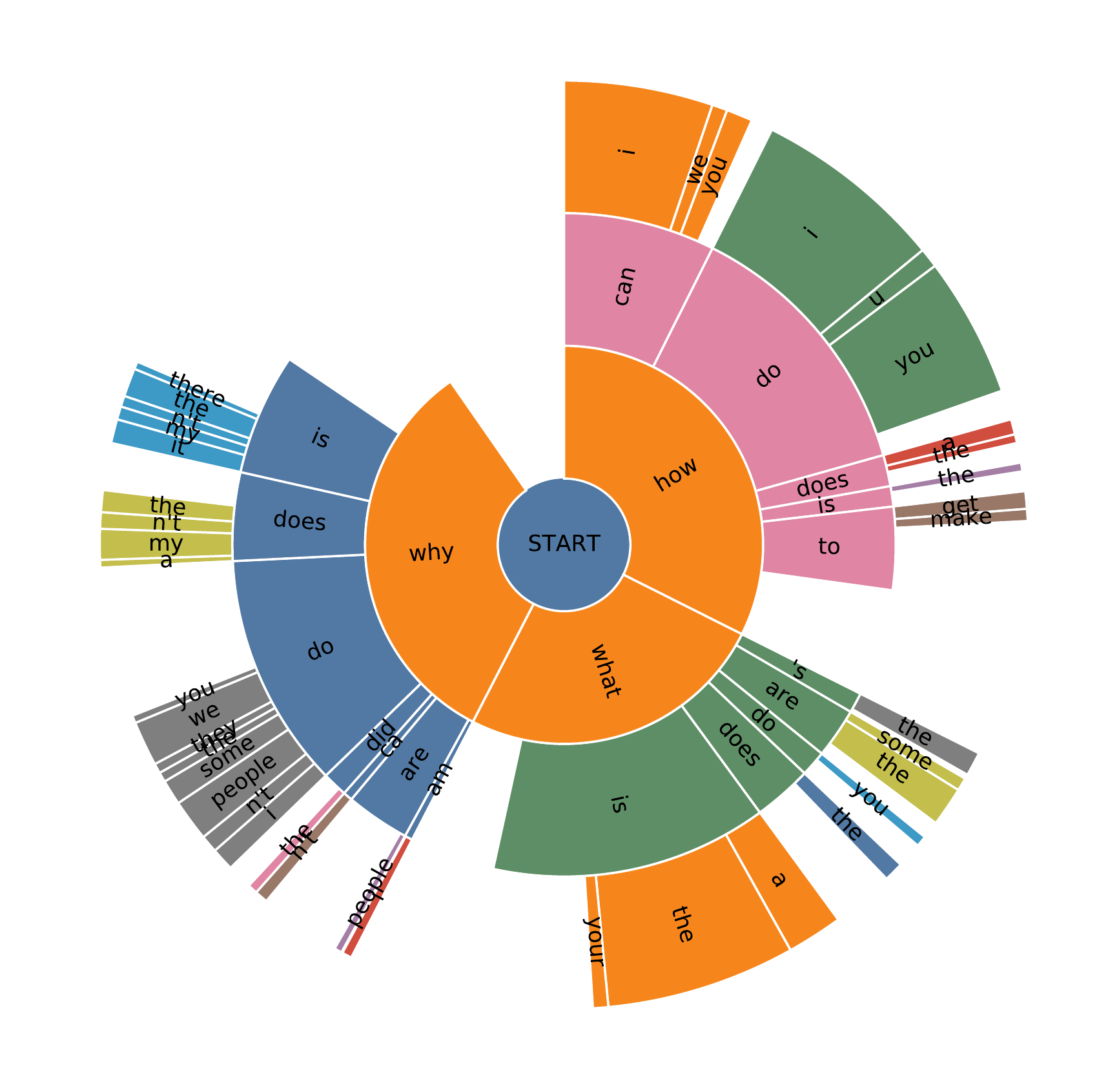}
    \vspace{-.9cm}
    \caption{Distribution of the top trigrams of \data questions (best viewed in color).}
    \label{fig:sunburst}
\end{figure}

\begin{table*}[t]
    \vspace{-0.4cm}
    \caption{The benchmark results by a wide variety of retrieval models on the \data dataset.}
    \vspace{-0.3cm}
    \centering
    \begin{tabular}{lcccccccc}\toprule
         \textbf{Method} & \textbf{MAP} & \textbf{MRR} & \textbf{P@1} & \textbf{P@3} & \textbf{P@10} & \textbf{nDCG@1} & \textbf{nDCG@3} & \textbf{nDCG@10} \\\midrule
         BM25 & 0.1977 & 0.4885 & 0.3333 & 0.2929 & 0.2485 & 0.4411 & 0.4237 & 0.4334\\
        
         DRMM-TKS~\cite{Guo:CIKM:2016} & 0.2315 & 0.5774 & 0.4337 & 0.3827 & 0.3005 & 0.4949 & 0.4626 & 0.4531\\
         aNMM~\cite{Yang:2016} & 0.2563 & 0.6250 & 0.4847 & 0.4388 & 0.3306 & 0.5289 & 0.5127 & 0.4904\\
         BERT~\cite{Devlin:2018} &\textbf{0.3771} & \textbf{0.7968} & \textbf{0.7092} & \textbf{0.6071} & \textbf{0.4791} & \textbf{0.7126} & \textbf{0.6570} & \textbf{0.6423}\\
         \bottomrule
    \end{tabular}
    \label{tab:results}
    \vspace{-0.2cm}
\end{table*}

\myparagraph{Training Set}
In the training set, we annotate the list $\mathcal{A}$ (see Section~\ref{sec:data}), for each query, and assume that for each question, answers to the other questions are irrelevant. 
As we removed similar questions from the dataset, this assumption is fair.
To test this assumption, we sampled 100 questions from the filtered version of nfL6 and annotated the top 10 results retrieved by BM25 using the same crowdsourcing procedure. The results showed that only 13.7\% of the documents were annotated as relevant (label 3 or 4). This error rate can be tolerated in the training process as it enables us to collect significantly larger amount of training labels. On the other hand, for the test set we performed pooling to label all possibly relevant answers. In total, the \data's training set contains 27,422 answer annotations as it shown in Table~\ref{tab:stats}, that is 11.3 annotated candidate answers per training question, which is significantly larger than its similar datasets, e.g., WikiPassageQA~\cite{Cohen:2018:SIGIR}.

\myparagraph{Test Set}
The test set in \data consists of 200 questions which were randomly sampled from nfL6 after pre-processing and filtering. Statistics of the test set can be found in Table\ref{tab:stats}. The set of candidate questions for annotation was selected by performing depth-$k$ ($k=10$) pooling. To do so, we considered the union of the top $k$ results of various retrieval models, including term-matching and neural models (listed in Table~\ref{tab:results}). We took the union of this set and 'nbest answers' (set $\mathcal{A}$) for annotation.

\vspace{-0.2cm}
\section{Data Analysis}
\label{sec:analysis}
In this section, we present a brief analysis of \data to highlight its characteristics. 

\myparagraph{Statistics of \data}
Table~\ref{tab:stats} lists general statistics of \data. As we see, \data consists of 2,426 non-factoid questions that can be used for training, followed by 200 questions as a test set. Furthermore, \data contains 27.4k and 6.5k annotations (judged answers) for the train and test sets, respectively. We also report the total number of answers with specific labels. 

\myparagraph{Workers Performance}
Overall, we launched 7 different crowdsourcing batches to collect \data. This allowed us to identify and ban less effective workers. As we see in Table~\ref{tab:stats}, a total number of 577 workers made over 148k annotations (257 per worker), out of which we rejected 12\% because they failed to satisfy the quality criteria.

\begin{figure}
    \centering
    \includegraphics[width=\columnwidth]{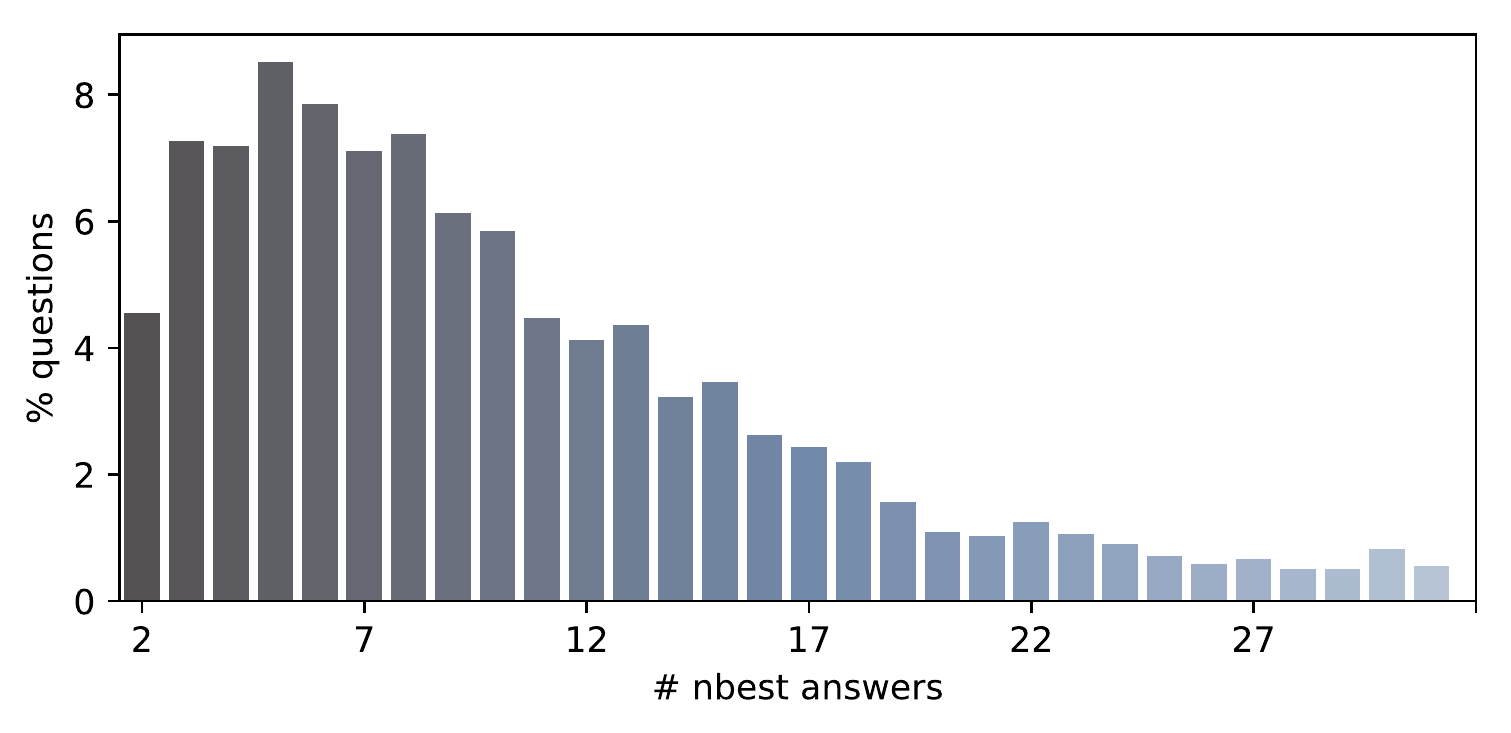}
    \vspace{-0.7cm}
    \caption{Distribution of the length of $\mathcal{A}$ (i.e., nbest answers) per question.}
    \label{fig:nbest}
    \vspace{-0.4cm}
\end{figure}


\myparagraph{Questions Distribution}
Figure~\ref{fig:sunburst} shows how questions are distributed in \data by reporting the top 40 starting trigrams of the questions. As shown in the figure, majority of the questions start with ``how'' and ``why,'' constituting 38\% and 36\% of the questions, respectively.
It is notable that, according to Figure~\ref{fig:sunburst}, a considerable number of questions start with ``how do you,'' ``how can you,'' ``what do you,'' and ``why do you,'' suggesting that their corresponding answers would be highly subjective and opinion based. Also, we can see a major fraction of questions start with ``how can I'' and ``how do I,'' indicating the importance and dominance of personal questions.


\myparagraph{Answers Distribution}
Finally, in Figure~\ref{fig:nbest}, we plot the distribution for the number of `nbest answers' ($|\mathcal{A}|$). We see that the majority of questions have 9 or less nbest answers (=54\%) and 82\% of questions have 14 or less nbest answers.
The distribution, however, has a long tail which is not shown in the figure. 
\section{Benchmark Results}
\label{sec:exp}

In this section, we provide benchmark results on the \data dataset. To this aim, we report the results for a wide range of retrieval models (mostly neural models) in Table~\ref{tab:results}. In this experiment, we report a wide range of standard retrieval metrics, ranging from precision- to recall-oriented metrics (see Table~\ref{tab:results}). Note that for the metrics that require binary labels (i.e., MAP, MRR, and P@k), we assume that the labels 3 and 4 are relevant, while 1 and 2 are non-relevant. Due to the definition of our labels (see Section~\ref{sec:data}), we recommend this setting for future work. For nDCG, we use the four-level relevance annotations.\footnote{Note that we mapped our 1 to 4 labels to 0 to 3 for computing nDCG.}

As shown in the table, the neural model significantly outperforms BM25, an effective term-matching retrieval model. Among all, BERT \cite{Devlin:2018} provides the best performance. Recent work on passage retrieval also made similar observations~\cite{Nogueira:2019,Padigela:2019}. Since MAP is a recall-oriented metric, the results suggest that all the models still fail at retrieving all relevant answers. There is still a large room for improvement, in terms of both precision- and recall-oriented metrics. 


\section{Conclusions}
In this paper, we introduced \data; a non-factoid community question answering dataset. The questions in \data were sampled from a wide range of categories on Yahoo! Answers, a community question answering service. We collected four-level relevance annotations through a multi-stage crowdsourcing as well as expert annotation. In summary, \data consists of 34,011 QA-pair relevance annotations for 2,426 and 200 questions in the training and test sets, respectively. Additionally, we reported the benchmark results for a set of retrieval models, ranging from term-matching to recent neural ranking models, on \data. Our data analysis and retrieval experiments demonstrated that \data introduces unique challenges while fostering research in the domain of non-factoid question answering. 

\section{Acknowledgement}
This work was supported in part by the Center for Intelligent Information Retrieval and in part by NSF IIS-1715095. Any opinions, findings and conclusions or recommendations expressed in this material are those of the authors and do not necessarily reflect those of the sponsor.



\end{document}